\documentstyle[editedvolume,numreferences]{crckapb}

\newcommand{\beq}{\begin{equation}}
\newcommand{\eeq}{\end{equation}}
\newcommand{\beqa}{\begin{eqnarray}}
\newcommand{\eeqa}{\end{eqnarray}}
\newcommand{\beqar}{\begin{eqnarray*}}
\newcommand{\eeqar}{\end{eqnarray*}}
\newcommand{\reef}[1]{(\ref{#1})}
\newcommand{\labell}[1]{\label{#1}} 
\newcommand{\ie}{{\it i.e.,}\ }
\newcommand{\eg}{{\it e.g.,}\ }
\newcommand{\ssc}{\scriptscriptstyle}
\def\S{{\cal S}}

\begin{opening}
\title{BLACK HOLES IN\protect\\
       HIGHER CURVATURE GRAVITY$^1$}

\author{R.C. Myers}
\institute{Department of Physics\\
           McGill University\\
           Montr\'eal, Qu\'ebec, Canada H3A 2T8}

\end{opening}

\runningtitle{HIGHER CURVATURE GRAVITY}

\begin{document}

\section{Introduction}
\addtocounter{footnote}{1}\footnotetext{To appear in:
{\it Black Holes, Gravitational Radiation, and the Universe:
Essays in Honor of C.V. Vishveshwara}, eds., C.V. Vishveshwara, B.R. Iyer,
B. Bhawal}

The idea that the Einstein action should be modified by the addition
of interactions
involving higher powers of the Riemann curvature tensor has
a long history stretching back to the early days of general relativity.
Such higher curvature theories
originally appeared in proposals by Weyl and Eddington
for a geometric unification
of electromagnetism and gravity\cite{early}. Much later, interest
arose in higher
derivative theories of gravity because
they provided renormalizable quantum field
theories\cite{renorma,renormb}. Unfortunately, the new massive
spin-two excitations, which tame the ultraviolet divergences in such
theories, result in the instability of the classical theory\cite{nege}
and the loss of unitarity in the quantum theory\cite{renormb,sad}.
While higher curvature theories have thus proven inadequate as
the foundation of quantum gravity,
they still have a role to play within the modern
paradigm of effective field theories\cite{win}.

Irrespective of the fundamental nature of 
quantum gravity, there should be a low energy effective
action which describes the dynamics of a
``background metric field''
for sufficiently weak curvatures and sufficiently long distances. On
general grounds, this effective gravity action will consist
of the usual Einstein action plus a series of
covariant, higher-dimension interactions,
{\it i.e.,} higher curvature terms, and also higher derivative terms
involving the ``low-energy'' matter fields. The appearance of such
interactions can be seen, for example,
in the renormalization of quantum field
theory in curved space-time\cite{field}, or in the construction of
low-energy effective actions for string theory\cite{string,wadia}.
In this context, the higher curvature interactions
simply produce benign perturbative corrections to
Einstein gravity coupled to conventional
matter fields\cite{nog,zzz}. 

Naive dimensional analysis would suggest that the coefficients of
the higher dimension terms in such an effective Lagrangian should
be dimensionless
numbers of order unity times the appropriate power of the Planck length.
Thus one might worry that all of the effects of the higher curvature terms
would be the same order as those of quantum fluctuations, and so there
would seem to be little point in studying modifications of
``classical'' black holes from higher dimension terms.
One motivation for
studying a classical higher curvature theory is that it is, of course,
possible that the coefficients of some higher dimension terms are larger
than what would be expected from simple dimensional analysis. 
Moreover, it is interesting to explore black holes in
generalized gravity theories in order to discover which properties
are peculiar to Einstein gravity, and which
are robust features of all generally covariant theories of gravity.
Further even within this framework, one may still discover clues
as to the ultimate nature of the underlying theory.

With these introductory remarks, we will go on to discuss some
of the recent investigations of black holes in higher curvature theories
of gravity. The remainder of this chapter is organized as follows: In
sect.~2, we briefly review various black hole solutions
which appear in the literature.
In sect.~3, we focus on black hole thermodynamics\cite{beker}
in the context of higher curvature gravity.
Finally in sect.~4, we provide a brief discussion and
indicate some of the open questions.

At this point, let us add that many of the recent candidates
for a theory of quantum gravity, especially those which attempt
to unify gravity with other interactions, are theories for which
the space-time dimension is greater than four.
Thus much of the following discussion refers to black holes
in higher dimensional space-times\cite{merry}.
While this idea may seem unusual and/or unappealing to some
readers, it is certainly one familiar to our esteemed colleague,
C.V. Vishveshwara\cite{visha,vishb}, to whom this volume is dedicated.
Throughout, we also employ the conventions of ref.~\cite{exam}
in any formulae.

\section{Black Hole Solutions}

When faced with the task of finding solutions of higher curvature
gravity (or any higher derivative theory), one must realize
that it is incorrect to assume that adding higher derivative
correction terms with ``small'' coefficients will {\it only}
produce small modifications of the solutions of the unperturbed
theory\cite{zzz}.
Any higher curvature theory will, in fact, contain whole classes
of new solutions unavailable to the classical Einstein theory.
In particular, the set of maximally symmetric vacuum solutions
will now include some number of (anti-)deSitter vacua, as
well as flat space\cite{round,desy}. A common feature of these
new solutions is that they are not analytic in the coefficients
of the higher derivative interactions, and hence in the context of
an effective field theory, they should be regarded as unphysical.
Systematic procedures have been developed to exclude these spurious
solutions\cite{zzz,lola}.
However, in the context of effective field theory where the domain
of validity
of the equations of motion is expected to be limited, it suffices
to treat the higher curvature contributions perturbatively.

For example, let us consider the field equations:
$R_{ab}=\alpha\,H_{ab}$ where $R_{ab}$ is the Ricci tensor, $H_{ab}$
is some higher derivative contribution, and $\alpha$ is the
(dimensionful) coefficient of the higher curvature term in the action.
Now making the obvious expansion of the solution,
$g_{ab}=g^0_{ab}+\alpha\,g^1_{ab}+\ldots$, one must solve
\beqa
R_{ab}(g^0)&=&0\labell{one}\\
\left(\Delta[g^0]\,g^1\right)_{ab}&=&H_{ab}(g^0)\labell{two}
\eeqa
where $\Delta[g^0]$ is the second order operator obtained from linearizing
the Ricci tensor about the metric $g^0_{ab}$. Solving eq.~\reef{one}
amounts to selecting a solution of the unperturbed Einstein theory.
Solving eq.~\reef{two} is solving for a linearized fluctuation in
that background with some source term. As well as imposing appropriate
boundary conditions on $g^1_{ab}$ (\eg that it preserve asymptotic
flatness and regularity of the horizon), one would (usually) choose
it to preserve the symmetries of the original solution (\eg
spherical symmetry, stationarity). While the above
discussion was phrased in terms of purely gravitational solutions,
it would be straightforward to include matter fields, as one might
in considering charged black holes. Further this approach extends in
an obvious way to developing the perturbation expansion to higher orders, 
which may involve including additional higher order interactions.

In this expansion,
the coupling constant $\alpha$ has the dimension of length to
some (positive) power $n$, \eg $n=2$ if $H_{ab}$ arises from a
curvature-squared interaction. The true dimensionless expansion
parameter in the above analysis is then $\alpha/L^n$ where $L$
is the local curvature scale. Hence this expansion will always
break down in the black hole interior near the singularity,
but it will be reliable in the exterior region of a large black hole
where the curvatures are small.
Thus within this
context, one can expect that for large black holes the asymptotic
regions feel only minor corrections
due to the higher curvature terms. Near the singularity,
the higher curvature contributions become strong, but this implies
that one has left the realm in which the effective action is to be
trusted. It would seem that at this point one must come to grips with
the full underlying fundamental theory. However, one might attempt to make
models of high curvature behavior\cite{frolov,morgan} to guide our
intuition.

Within this framework, it is interesting to consider the propagation
of metric disturbances in the black hole background. One would
organize the disturbance with the same $\alpha$ expansion as
above: $g_{ab}=(g^B+h)_{ab}=(g^0+h^0)_{ab}+\alpha(g^1+h^1)_{ab}$
where $g^B_{ab}=g^0_{ab}+\alpha\,g^1_{ab}$ is the background metric
satisfying eqs.~\reef{one} and \reef{two}, above. The disturbance then
satisfies
\beqa
\left(\Delta[g^0]\,h^0\right)_{ab}&=&0\labell{three}\\
\left(\Delta[g^0]\,h^1\right)_{ab}&=&J_{ab}(g^0,h^0)\labell{four}
\eeqa
where $J_{ab}(g^0,h^0)$ is the linearization of $H_{ab}(g^0+h^0)$.
Note that within this scheme, the propagation of metric disturbances
and hence the causal structure are completely determined by the
original background metric $g^0_{ab}$. Hence we are guaranteed that
the black hole really remains a black hole, and the event
horizon remains an event horizon. 
 
At some level, the conclusions of the previous paragraph
may actually seem somewhat surprising. For instance, one would
conclude that rather than following null geodesics in the perturbed
background $g^B_{ab}=g^0_{ab}+\alpha\,g^1_{ab}$, ``high frequency''
gravity waves still follow null geodesics of the original metric
$g^0_{ab}$. In either case, one would expect these conclusions to apply
in an approximation: $L^n>>\lambda^n>>\alpha$ where $\lambda$ is the
wavelength of the disturbance. The first inequality justifies a
geometric optics approximation\cite{exam}, whereas the second
is required for the reliability of the effective action. Now in
examining the propagation of wavefronts in curved space, one
expects curvature corrections to appear
at the order $\lambda/L$, whereas the modifications to the metric
due to the higher curvature interactions are of the order $\alpha/L^n$.
Hence given the above inequalities, one  has $\lambda/L>>(\lambda/L)^n
>>\alpha/L^n$, and so the discrepancy between using
either $g^B_{ab}$ or $g^0_{ab}$ to define null geodesics is certainly
a subleading correction in determining the propagation of gravitational
disturbances. However, using $g^0_{ab}$ seems more
consistent in the application of perturbation theory\cite{horror,faster}.

Eqs.~\reef{three} and \reef{four}
are also relevant in addressing the important question of the (linearized)
stability of the horizon\cite{stable} --- a topic to which Vishu
has made seminal contributions\cite{vishc}. Given that
the original Einstein theory was stable, one knows that
there are no runaway solutions to eq.~\reef{three}. Eq.~\reef{four}
simply extends the latter equation by the addition of a regular
source term. Hence it is also clear that $h^1_{ab}$ has no runaway
solutions by the application of the original stability
analysis. Thus one can immediately deduce that the black hole
solution remains stable within this perturbative framework.

A perturbative approach has been applied in examining
modifications of the four-dimensional Schwarzschild black hole within the
context of renormalized Einstein gravity\cite{perta,pertb}.
In this case, curvature-squared interactions do not produce
any modifications for solutions of the four-dimensional
vacuum Einstein equations. Hence this analysis considered
Einstein gravity perturbed by the addition of terms involving
three Riemann curvatures. The main observation
resulting from this analysis\cite{perta} was that the relations between
the black hole's
mass and its thermodynamic parameters\cite{beker} are modified.
In particular, the black hole entropy was not longer proportional to
the area of the horizon.

Various perturbative analyses have also been made to study black
holes in string theory\cite{stb}--\cite{std}. These include
considering the effects of curvature-squared terms on spherically
symmetric black holes in arbitrary dimensions\cite{sta}, and on
four-dimensional black holes with angular momentum\cite{stc}
or with charge\cite{std}.\footnote{Note that in string theory,
charged black holes typically differ from the Reissner-Nordstrom
geometry as a result of nonminimal couplings between the
matter fields\cite{stcharge}.}
In certain supersymmetric string theories, the
leading higher curvature interaction can be shown to contain
four curvatures\cite{four}.
The effect of these terms on spherically symmetric black holes
in arbitrary dimensions has also been studied\cite{stb}.
Apart from modifications to the usual thermodynamic properties of the
black holes, one of the remarkable observations here was the fact that
the higher curvature terms induce various new forms of long-range
scalar field hair on the black holes. However, this new hair
may be regarded as secondary\cite{coal}, in that it is
completely determined by the black hole's primary hair,
{\it e.g.,} the mass and charge. In other words,
there are no new constants of integration, and hence
these solutions do not violate the spirit of the no-hair
theorems\cite{hair,hare}.
Essentially, the new hair arises because
the scalar fields have non-minimal couplings to the
higher curvature terms. 

Motivated originally by string theory, a great deal of attention
has been focussed on Lovelock gravity\cite{love}. The latter is defined
by a Lagrangian which
is the sum of dimensionally extended Euler densities.
In four dimensions, all of the higher curvature terms are
total derivatives, and hence the theory reduces to Einstein gravity.
However, in higher dimensions, the new interactions do make
nontrivial contributions. A distinguishing feature of
these Lagrangians is that the resulting equations
of motion contain no more than second derivatives in time\cite{love}.
As quantum theories then, they
are free of ghosts when expanding about flat space\cite{zwi},
and so they evade the problem of unitarity loss ---
however, they remain
nonrenormalizable. Exact spherically symmetric solutions were first
found for the curvature-squared or Gauss-Bonnet
theory\cite{desy}. These results were
quickly extended to arbitrary Lovelock theories\cite{jt}--\cite{also}, as well
as charged black holes\cite{wilt}. These solutions
displayed a rich structure of multiple horizons, and unusual
thermodynamic properties\cite{right}. For example, certain (uncharged)
solutions could be found with vanishing Hawking temperature\cite{righto}.
The topic of Lovelock black holes is, in fact, another area upon
which Vishu's research has touched\cite{vishb}.

While in four dimensions, the Lovelock action yields only Einstein
gravity, one can also consider the Kaluza-Klein compactification\cite{kk}
of a higher dimensional Lovelock theory down to four dimensions.
The resulting
theory consists of Einstein gravity coupled to 
various scalar and vector fields with nonminimal higher-derivative
interactions\cite{folk}. In this case, there are four-dimensional black holes
which carry secondary scalar hair\cite{eric}.

More recently researchers\cite{fulla,fullb}, interested in whether the 
secondary scalar
hair found in ref.~\cite{sta}--\cite{std} survived beyond perturbation theory,
investigated exact solutions of a (four-dimensional)
dilatonic Gauss-Bonnet theory. In the latter, the curvature-squared
interaction is modified by the addition of
a nonminimal scalar coupling.
The full equations of this theory are difficult enough that
they could not be solved analytically. However, analytic arguments
and numerical evidence indicates that the modified black holes do
carry secondary scalar hair\cite{fulla}. These results were also
extended to black holes carrying charge\cite{fullb}.

Some work\cite{pola,polb}
 has also been done on black holes in theories where the
Lagrangian density takes the form $\sqrt{-g}f(R)$ where $f(R)$ is
a polynomial in the Ricci scalar. These models are
amenable to analysis because they can be mapped to a theory
of Einstein gravity and a minimally coupled scalar with an unusual
potential\cite{pola,polc}. Note that if other matter fields
are included then the latter develop unusual nonminimal couplings
in the Einstein-scalar theory. In the case $f=R+a_2R^2$,
a uniqueness theorem was proven for certain classes of matter
fields in four dimensions\cite{pola}.
Provided\footnote{Note that this condition
is also required for the stability of the theory\cite{poland,white}.}
that $a_2>0$, no new hair can arise and
so the only black hole solutions are identical to those of Einstein gravity.
Ref.~\cite{polb} extended this work to spherically symmetric
solutions for general
polynomial actions in arbitrary dimensions.
One finds that, with the same restriction on the quadratic term and
irrespective of the remaining terms in the polynomial $f$,
the only asymptotically flat black holes are still the
Schwarzschild solutions.

The latter investigations of the Lovelock, dilatonic Gauss-Bonnet and
polynomial-in-$R$ theories all go beyond the perturbative approach
originally described and consider exact solutions of the full higher
derivative equations of motion. Certainly amongst the exact solutions,
one will find some which lend themselves to a Taylor expansion in the
higher curvature coupling constants. These solutions could then be considered
the result of carrying out the perturbation expansion to infinite order.
To be of interest in the effective field theory framework though, one
would have to know that there are no additional higher
curvature interactions at higher orders. In general, this seems an
unlikely scenario, and in string theory, certainly one that does
not apply\cite{john}\footnote{In certain cases, however, one can
use supersymmetry to argue that special solutions are exact to all
orders despite the higher curvature corrections to the
action\cite{exact}.} Although the physical motivation may not be strong,
one can still set out to study these theories and their solutions
as a mathematical problem in its  own right. In this respect, a common
advantageous feature, which the three theories discussed here seem to
share in common, is
the absence of negative energy ghosts\cite{zwi,poland,white,desb}
 --- at least with certain restrictions on the coupling constants.

{}From this point of view, one should readdress the important question
of the stability of the horizon for black hole solutions in these theories.
Of course, the linearized stability for
the full higher derivative equations is an extremely
difficult problem\footnote{Note that going to higher dimensions introduces
complications by itself\cite{ray}.},
however, some limited results have been achieved.
 Ref.~\cite{white} shows that the four-dimensional
Schwarzschild black hole is stable in a general fourth order gravity theory.
There also some limited results indicating stability of spherically
symmetric four-dimensional black holes in the dilatonic Gauss-Bonnet
theory\cite{lineal}.

A more fundamental question which should also be considered
is the actual causal structure of the ``black hole'' solutions.
It would seem that for the three theories of interest here, the
Lovelock, dilatonic Gauss-Bonnet and polynomial-in-$R$ gravities,
that the full equations are (or can be mapped to) second-order
hyperbolic systems. However, 
the characteristic surfaces of these equations need not coincide
with null cones in the background space-time, opening the
possibility that gravitational disturbances could propagate
``faster than light.'' Note that such acausal behavior would
result immediately if the theory in question had negative
energy ghosts\cite{acausal}, which, although a feature of generic
higher curvature theories, is not the case for these three theories.
The possibility that gravitational disturbances could escape
a ``black hole'' would have profound consequences for these
theories. In the case of the Lovelock theories, investigations
have been made of the modified characteristics\cite{char}, and
in this case, a preliminary
study\cite{abort} of radial wavefronts in static black holes suggests 
that disturbances can not escape the horizon.
The same result would necessarily seem to apply for the polynomial-in-$R$
theories, which are essentially mapped to Einstein gravity (with a
conformally related metric) coupled to a massive scalar field\cite{polc}.
We stress, however, that we view the motivation in studying these theories
as more mathematical than physical.

\section{Black Hole Thermodynamics}

Black hole thermodynamics\cite{beker} is certainly one of the most
remarkable features of the area of physics to which this volume is
devoted. It produces a confluence of ideas from thermodynamics,
quantum field theory and general relativity. Much of the interest
in black hole thermodynamics  comes from the hope that it will provide
some insight into the nature of quantum gravity. While originally
developed in the context of Einstein gravity, it is easy to see
that much of the framework should extend to higher curvature theories,
as well. Hawking's celebrated result\cite{hawk} that a black hole emits
thermal radiation with a temperature proportional to its surface
gravity, $\kappa$:
\beq
k_{\ssc B}T={\hbar\kappa\over 2\pi c}
\labell{hawkt}
\eeq
is a prediction of quantum field theory in space-time containing
a horizon\cite{field}, independent of the details of the dynamics
of the gravity theory. Alternatively, this result will follow from the
evaluation of the of the black hole partition function using the
Euclidean path integral method\cite{euclid}.
Applying the latter approach in higher curvature gravity, it is also clear that 
one can derive a version of the First
Law of black hole mechanics
\beq
{\kappa\over2\pi c}\delta\S= c^2\delta M-\Omega^{(\alpha)}\delta J_{(\alpha)}
\labell{firstl}
\eeq
where $M$, $J_{(\alpha)}$ and $\Omega^{(\alpha)}$ are the black hole mass,
canonical angular momentum, and the angular velocity of the
horizon\cite{explic}. Given the identification of the black hole
temperature with the surface gravity in eq.~\reef{hawkt}, the black
hole entropy is naturally identified as $S=(k_{\ssc B}/\hbar)\S$. In
the context of Einstein gravity, this gives the famous
Bekenstein-Hawking entropy\cite{hawk,bek}:
\beq
S_{\ssc BH}={k_{\ssc B}c^3\over\hbar G} {A_{\ssc H}\over 4}\ .
\labell{bhent}
\eeq
where $A_{\ssc H}$ is the area of the event horizon\cite{exp}.

Many of the early
investigations\cite{perta,sta,stb,right,wilt,righto}
which examined particular black hole solutions
in higher curvature theories noted that eq.~\reef{bhent}
no longer applied --- see ref.~\cite{failS} for a review.
Important conceptual progress was made in ref.~\cite{lovely},
where it was realized that $\S$ should take the form of
a geometric expression evaluated at the event horizon. This result
was explicitly demonstrated there for the special case of Lovelock gravity,
by extending a Hamiltonian derivation of the First Law\cite{suds}.
Various other techniques were then developed\cite{wald1}--\cite{late}
to expand this result to other higher curvature theories.
In particular though,
Wald\cite{wald1} developed an elegant new derivation of the First Law
which applies for any
diffeomorphism invariant theory --- see below.
His derivation makes clear that in eq.~\reef{firstl}, $\S$ may always
be expressed as a {\it local} geometric density
integrated over a space-like cross-section of the horizon.

\subsection{Black Hole Entropy as Noether Charge}

Here, we will provide a brief introduction to Wald's derivation
of the First Law.
The interested reader is referred to Refs.~\cite{wald1,wald2,onS,waldrev}
for a complete description.
In the following, we also adopt the standard
convention of setting $\hbar=c=k_{\scriptscriptstyle B}=1$.

An essential element of Wald's approach is the
Noether current associated with diffeomorphisms\cite{wallee}.
Let $L$ be a Lagrangian built out of some set of
dynamical fields,
including the metric, collectively denoted as $\psi$.
Under a general field variation $\delta \psi$, the Lagrangian
varies as
\begin{equation}
\delta(\sqrt{-g} L)=\sqrt{-g}E\cdot\delta \psi + \sqrt{-g}\,\nabla_{\! a}
\theta^a(\delta \psi)\ \ ,
\label{dL}
\end{equation}
where ``$\cdot$" denotes a summation over the dynamical
fields including contractions of tensor indices.
Then the equations of motion are $E=0$.
With symmetry variations for which $\delta(\sqrt{-g} L)
=0$, $\theta^a$ is the Noether current which is conserved
when the equations of motion are satisfied --- {\it i.e.,} $\nabla_{\! a}
\theta^a(\delta \psi)=0$ when $E=0$. 
For diffeomorphisms, where the field variations are given by the
Lie derivative $\delta\psi={\cal L}_\xi\psi$, 
the variation of a covariant Lagrangian is a total derivative,
$\delta (\sqrt{-g} L)={\cal L}_\xi(\sqrt{-g} L)
=\sqrt{-g}\nabla_{\! a}(\xi^a\, L)$.
Thus one constructs an improved Noether current,
\[
J^a=\theta^a({\cal L}_\xi \psi)- \xi^a L\ \ ,
\]
which satisfies $\nabla_{\!a}J^a=0$ when $E=0$.

A fact\cite{wald3}, which may not be well-appreciated, is that for {\it any}
 local symmetry,
there exists  a globally-defined Noether potential $Q^{ab}$,
satisfying $J^a=\nabla_{\! b}Q^{ab}$
where $Q^{ab}=-Q^{ba}$.
$Q^{ab}$ is a local function of the dynamical fields and a
linear function of the symmetry parameter ({\it i.e.,} $\xi^a$
in the present case). Of course,
this equation for $J^a$ is valid up to terms which vanish
when the equations of motion are satisfied.
Given this expression for $J^a$,
it follows that the Noether charge contained in a spatial volume $\Sigma$
can be expressed as a boundary integral $\oint_{\partial \Sigma}d^{D-2}\!x\,
\sqrt{h}\,\epsilon_{ab}Q^{ab}$, where $h_{ab}$ and
$\epsilon_{ab}$ are the induced metric and binormal form
on the boundary $\partial \Sigma$.

Another key concept that enters in Wald's construction is that
of a Killing horizon.
Given a Killing vector field which generates an
invariance for a particular solution --- {\it i.e.,} ${\cal L}_\xi\psi=0$
for all fields --- a Killing horizon is
a null hypersurface whose null
generators are orbits of the Killing vector.
If the horizon generators are geodesically
complete to the past (and if the surface gravity is nonvanishing),
then the Killing horizon
contains a space-like cross-section $B$, the {\it bifurcation surface},
on which the Killing field $\chi^a$
vanishes\cite{raczwald}. It can be shown that the event horizon
of any black hole, which is static or is stationary with a certain
$t$--$\phi^{(\alpha)}$ orthogonality condition, must be a Killing
horizon\cite{card}.\footnote{Note
that the Zeroth Law, \ie the constancy of of the surface gravity
over a stationary event horizon, follows if the latter is also
a Killing horizon\cite{card,rack}. This is significant since
the Zeroth Law is actually an
essential ingredient to the entire framework of black hole
thermodynamics.}
A stronger result holds in general relativity,
where it can be proven that the event horizon for any stationary
black hole is Killing\cite{area}.

The key to Wald's derivation of the First Law is the identity
\beq
\delta H=\delta\int_\Sigma dV_a J^a -
\int_\Sigma dV_a\nabla_b(\xi^{a}\theta^{b}-\xi^{b}\theta^{a} ),
\labell{clef}
\eeq
where $H$ is the Hamiltonian generating evolution along
the vector field $\xi^a$, and
$\Sigma$ is a spatial hypersurface with volume element $dV_a$.
This identity is satisfied for arbitrary variations of the fields
away from any background solution.
If the variation is to another solution, then one can replace
$J^a$ by $\nabla_b Q^{ab}$, so the variation of the Hamiltonian
is given by surface integrals over the boundary $\partial\Sigma$. 
Further, if $\xi^a$ is a Killing vector of the background solution, then
$\delta H=0$, and in this case, one obtains an identity relating the
various surface integrals over $\partial\Sigma$.

Suppose that the background solution is chosen to be a
stationary black hole with horizon-generating Killing field
$\chi^a\partial_a={\partial_t}+\Omega^{(\alpha)}
{\partial_{\phi^{(\alpha)}}}$,
and the hypersurface $\Sigma$ is chosen to
extend from asymptotic infinity down
to the bifurcation surface where $\chi^a$ vanishes.
The surface integrals at infinity then yield precisely the
mass and angular momentum variations, $\delta M-\Omega^{(\alpha)}
\delta J_{(\alpha)}$, appearing in Eq.~(\ref{firstl}),
while the surface integral at the bifurcation surface reduces to
$\delta\oint_Bd^{D-2}\!x\,\sqrt{h}\,\epsilon_{ab}Q^{ab}({\chi})$.
Finally, it can be shown that the latter surface integral always
has the form $(\kappa/2\pi) \delta \S$, where
$\kappa$ is the surface gravity of the background black hole, and
$\S=2\pi\oint_Bd^{D-2}\!x\,\sqrt{h}\,\epsilon_{ab} Q^{ab}({\tilde{\chi}})$,
with $\tilde{\chi}^a$, the Killing vector scaled to have unit
surface gravity.

By construction $Q^{ab}$ involves the Killing
field $\tilde{\chi}^a$ and its derivatives. However, this
dependence can be eliminated as follows\cite{wald1}: Using
Killing vector identities, $Q^{ab}$ becomes a function of only $\tilde{\chi}^a$
and the first derivative, $\nabla_{\! a}\tilde{\chi}_b$.
At the bifurcation surface, though, $\tilde{\chi}^a$ vanishes and
$\nabla_{\! a}\tilde{\chi}_b={\epsilon}_{ab}$, where ${\epsilon}_{ab}$
is the binormal to the bifurcation surface.
Thus, eliminating the term linear in $\tilde{\chi}^a$ and replacing
$\nabla_{\! a}\tilde{\chi}_b$ by ${\epsilon}_{ab}$ yields a completely
geometric
functional of the metric and the matter fields, which may be denoted
$\tilde{Q}^{ab}$. One can show that the
resulting expression,
\begin{equation}
\S=2\pi\oint d^{D-2}\!x\,\sqrt{h}\,\epsilon_{ab}\, \tilde{Q}^{ab},
\label{Sgeom}
\end{equation}
yields the
correct value for $\S$ when evaluated not only at the
bifurcation surface, but in fact on
an arbitrary cross-section of the Killing horizon\cite{wald2,onS}.

Using Wald's technique,
the formula for black hole entropy has been found
for a general Lagrangian of the following form:
\[
L=L(g_{ab}, R_{abcd}, \nabla_e R_{abcd},
\nabla_{(e_1}\nabla_{e_2)} R_{abcd}, \ldots;
\psi,\nabla_{\! a}\psi,\nabla_{(a_1}\nabla_{a_2)}\psi, \ldots)\ \ ,
\]
involving the Riemann tensor and symmetric derivatives
of $R_{abcd}$ (and the matter fields, denoted by $\psi$)
up to some finite order $n$. $\S$ may then be
written\cite{visser}--\cite{onS}
\beq
\S=-2\pi\oint d^2\!x\sqrt{h}\ \sum^n_{m=0} (-)^m\
\nabla_{(e_1}\ldots\nabla_{e_m)}Z^{e_1\cdots e_m:abcd}\
\epsilon_{ab}\,{\epsilon}_{cd}
\label{wform}
\eeq
where the $Z$-tensors are defined by
\beq
Z^{e_1\cdots e_m:abcd}\equiv {\delta\, {L}\over
\delta\nabla_{(e_1}\ldots\nabla_{e_m)} R_{abcd}}\ \ .
\eeq
As a more explicit example, consider a polynomial-in-$R$ action
\beq
I={1\over16\pi G}\int d^4\!x\,\sqrt{-g}\left(R+a_2R^2+a_3 R^3
\right)
\eeq
for which one finds the simple result
\beq
\S={1\over4G}\oint d^{D-2}\!x\,\sqrt{h}\left(1+2a_2 R+3a_3 R^2
\right)\ \ .
\label{resultsa}
\eeq
Similarly in the Gauss-Bonnet theory with action
\beq
I={1\over16\pi G}\int d^4\!x\,\sqrt{-g}\left(R+\alpha(R_{abcd}R^{abcd}
-4R_{ab}R^{ab}+R^2)
\right)
\eeq
one finds from eq.~\reef{wform} that
\beq
\S={1\over4G}\oint d^{D-2}\!x\,\sqrt{h}\left(1+2\alpha\, \tilde{R}(h)
\right)\ \ .
\label{resultsb}
\eeq
where $\tilde{R}(h)$ is the Ricci scalar calculated for the induced
metric $h_{ab}$. In both eqs.~\reef{resultsa} and \reef{resultsb},
the first term yields the expected contribution for Einstein
gravity, namely $A/(4G)$. Thus just as the Einstein term in the action
is corrected by higher-curvature terms, the Einstein
contribution to the black hole entropy receives higher-curvature corrections.

\section{Epilogue}

While higher curvature theories are typically pathological when considered
as fundamental, they may still be studied within the framework
of effective field theory\cite{win} where they produce minor corrections
to Einstein gravity. Still perturbative investigations\cite{perta}--\cite{std}
of black holes in this context
have revealed modifications to black hole thermodynamics and the
generation of new scalar hair in these theories. It may be of interest
to the perturbative framework on a more formal footing, \eg addressing
questions such as whether or not solutions to eq.~\reef{two} always
exist which leave the event horizon a regular surface.

The absence of ghosts in the Lovelock, dilatonic Gauss-Bonnet and
polynomial-in-$R$ theories also seems an interesting question to
investigate more fully. This feature seems to make these theories
an interesting mathematical framework in which to study exact black hole
solutions of the full higher derivative equations. It may be of interest
then to find more general stationary solutions, \ie rotating black hole
solutions. The stability of the event horizon, however, remains an
important question to be addressed for even the solutions with
spherical symmetry.

Ultimately it was the studies of various black holes solutions that
led to the beautiful generalization of black hole thermodynamics
for higher curvature theories. Wald's new derivation of the
First Law\cite{wald1} demonstrates the black hole entropy is always
determined by a {local} geometric expression evaluated at
the event horizon, and provides a general and explicit formula
\reef{wform} for $\S$. It may be that these results may be
used to provide an even more refined test of our understanding
of black hole entropy in superstring theory, given the recent dramatic
progress in that area\cite{wadia}. One interesting question related
to Wald's derivation is determining
the minimal requirements for a stationary event horizon to
be a Killing horizon.

While eq.~\reef{wform} provides an elegant expression for $\S$, it should
be noted that amongst the details overlooked in sect.~3.1 was the
fact that a number of ambiguities arise in the construction
of $\tilde{Q}^{ab}$ \cite{wald2,onS}. Hence eq.~\reef{wform} should be
understood as the result of making certain (natural) choices in
the calculation. None of these ambiguities have any
effect when $\S$ is evaluated on a stationary horizon, but they
might become significant for non-stationary horizons. As an example,
it should be noted that for the Gauss-Bonnet theory, eqs.~\reef{wform}
and \reef{resultsb} are strictly not the same because the latter expression
relies on the additional information that the extrinsic curvature of the
bifurcation surface vanishes\cite{lovely}. Hence the two expressions
should not be expected to coincide on a non-stationary horizon.

An important guide to resolving these ambiguities should be the
Second Law. Certainly if $\S$ is to play the role of an entropy,
it should also satisfy the Second Law of black hole thermodynamics
as a black hole evolves, \ie there should be a classical increase
theorem for any dynamical processes. Within Einstein gravity, the
Second Law is established by Hawking's area theorem\cite{area}.
So far only limited results have been produced for higher curvature
theories\cite{increase} --- see also comments in ref.~\cite{waldrev}.
One can show for quasistationary processes
that the Second Law is in fact a direct consequence of the First
Law (and a local positive energy condition for the matter fields),
independent of the details of the gravitational dynamics. For
polynomial-in-$R$ theories, one can establish the Second Law
with certain restrictions on the coupling constants (and again
a positive energy condition on the matter sector). One proof 
of the latter involves studying the properties of the
null rays along the event horizon with an extension of the
Raychaudhuri equation. A valuable extension of these results\cite{increase}
would be to establish the Second Law within a perturbative
framework. Of course, another important question is to determine
the validity of the generalized Second Law\cite{gsl} for evaporating black
holes in the higher curvature theories.

A final question which I pose here is related to the conformal
anomaly\cite{conform}. In quantum field theory\cite{field}, the conformal
anomaly may be characterized as higher derivative modifications of
the renormalized gravitational equations of motion which
do not result from the variation of a local action. It appears
that these terms will modify the expression for the black hole
entropy\cite{conforms} but they can not in general be addressed with Wald's
construction --- see, however, \cite{conform2}. 
It would be interesting to improve on the latter
to systematically include the effects of these contributions, and
to determine
whether $\S$ or its variation in the First Law still retains its
local geometric character. The investigation of a two-dimensional
toy model would seem to indicate that the answer to the last question
is negative\cite{conform2}.

\vskip 1cm
I would like to gratefully acknowledge useful discussions with
T.~Jacobson. I am also grateful for the hospitality of
the Institute for Theoretical Physics at the University of
California, Santa Barbara, where this manuscript was completed.
This work was supported by NSERC of Canada, and at the ITP, UCSB
by NSF Grant~PHY94--07194.


\begin{thebibliography}{99}
\bibitem[faster]
{faster}{We would advocate that the same reasoning should be applied
in the recent studies which suggest that as a result of nonminimal
couplings to curvature in an effective action, photon propagation
in curved space backgrounds can be ``superluminal''\cite{fast}.
One should not ascribe any physical
significance to the results derived there without a thorough investigation
of the expected curvature corrections to the propagation of
wave packets\cite{notso}.}
\bibitem[explic]
{explic}{In higher dimensions, one has the possibility of commuting
rotations in totally orthogonal planes. Hence the black hole is
characterized by the set of angular momenta $J_{(\alpha)}$ in the
maximal set of orthogonal planes. See ref.~\cite{merry}.}
\bibitem[exp]
{exp}{Note that in $D$ dimensions, a space-like cross-section of the horizon
is a ($D$--2)-dimensional space, and so the surface area $A_{\ssc H}$
refers to the volume of such a slice.}
\end{thebibliography}
\end{document}